\documentclass[12pt]{article}
\usepackage{amsbsy}
\usepackage{amsfonts}

\begin{document}

\def\theequation{\arabic{section}.\arabic{equation}}

\renewcommand\vec[1]{\boldsymbol{#1}}
\def\vzeta{\vec{\zeta}}
\def\vomega{\vec{\omega}}
\def\vdelta{\vec{\delta}}

\def\th#1#2#3{
   \theta^{\scriptscriptstyle #1}_{\scriptscriptstyle #2} \!
   \left( #3 \right) \, }

\def\pf#1#2{
   \hat\theta^{\scriptscriptstyle #1}_{\scriptscriptstyle #2}\, }

\title{Finite genus solutions for the Ablowitz-Ladik hierarchy.}
\author{V.E.Vekslerchik}
\maketitle

\begin{abstract}
The question of constructing the finite genus quasiperiodic solutions
for the Ablo\-witz-Ladik hierarchy (ALH) is considered by establishing
relations between the ALH and the Fay's identity for the
$\theta$-functions. It is shown that using a limiting procedure one can
derive from the latter an infinite number of differential identities which
can be arranged as an infinite set of differential-difference equations
coinciding with the equations of the ALH, and that the original Fay's
identity can be rewritten in a form similar to the functional equations
representing the ALH which have been derived in the previous works of the
author. This provides an algorithm for obtaining some class of
quasiperiodic solutions for the ALH, which can be viewed as an
alternative to the inverse scattering transform or the algebro-geometrical
approach.  
\end{abstract}

%%%%%%%%%%%%%%%%%%%%%%%%%%%%%%%%%%%%%%%%%%%%%%%%%%%%%%
\section{Introduction.}              \label{sec_Intro}
%%%%%%%%%%%%%%%%%%%%%%%%%%%%%%%%%%%%%%%%%%%%%%%%%%%%%%
\setcounter{equation}{0}

The problem of constructing the quasiperiodic solutions (QPS) is one of
the most challenging problems of the theory of integrable systems, and many
mathematicians and physicists spent much efforts to obtain the QPS for
almost all equations that are known to be integrable. The Ablowitz-Ladik
hierarchy (ALH), which has been introduced in \cite{AL1}, is not an
exception.  So, e.g, one should mention the works by N.N.Bogolyubov (Jr) 
et al \cite{BPS,BP} and S.Ahmad, A.R.Chowdhury \cite{AC1,AC2} 
devoted to the discrete nonlinear Schrodinger equation (DNLSE) and the
discrete modified Korteveg-de Vries equation (DMKdV), which are the best
studied equations of the ALH. There authors were studying this problem in
the framework of the inverse scattering transform (IST).  Another, the
so-called algebraic-geometrical, approach has been used by Miller,
Ercolani, Krichever and Levermore, who considered in \cite{MEKL} the
complex version of the DNLSE and obtained the Baker-Akhiezer function and
QPS corresponding to finite genus Riemann surfaces. This work provides
almost exhaustive solution of the problem of the finite genus QPS, but its
results need some further simplification to be useful for practical
purposes, especially if one wants to extend them to the higher equations
of the ALH and in this work I will try to avoid the algebro-geometrical
language, and will use some more direct (and simpler) strategy. As it has
been established in \cite{MEKL}, each finite genus QPS of the DNLSE can be
presented as a quotient of the $\theta$-functions of some arguments
multiplied by an exponent of some phase, all of them being some {\it
linear} functions of the coordinates (the same is true in the cases of the
DMKdV as well as all other equations of the hierarchy). Thus, since we
know the structure of the solutions, all we have to do to derive them is
to calculate some number of constant parameters. So, it desirable to
develop some method which will enable us to obtain these constants (and
hence solutions) straightforwardly, not using technique (sometimes rather
complicated) of the theory of the functions and differentials on
hyperelliptic Riemann surfaces. It turns out that this can be done.
Moreover, this can be done not only for the DNLSE or DMKdV but, in
principle, for all equations of the hierarchy simultaneously. Namely this
is the main question of this paper.  The key moment is that the
$\theta$-functions of the finite genus Riemann surfaces (of which the
finite genus QPS are built-up) satisfy some algebraic relation,
the so-called Fay's trisecant formula \cite{Fay,Mumford}, which can be
used to obtain an infinite number of differential identities, which, as
will be shown below, are closely related to the ALH, and can be used to
obtain the QPS we are looking for.  Such approach also demonstrates some
new, to my knowledge, feature of the ALH (and namely this was one of the
main motives to write this paper): the equations of the ALH naturally
appear when flows over Riemann surfaces are considered (I will return to
this question below).

The plan of the paper is as follows. After presenting some basic
facts on the ALH (section \ref{sec_ALH}) I will discuss the Fay's
formula and its differential consequences (section \ref{sec_Fay}).
These results will be used to obtain the finite genus QPS for the ALH
(section \ref{sec_QPS}).

%%%%%%%%%%%%%%%%%%%%%%%%%%%%%%%%%%%%%%%%%%%%%%%%%%%%%%
\section{Ablowitz-Ladik hierarchy}     \label{sec_ALH}
%%%%%%%%%%%%%%%%%%%%%%%%%%%%%%%%%%%%%%%%%%%%%%%%%%%%%%
\setcounter{equation}{0}

The ALH is an infinite set of integrable differential-difference
equations, which has been introduced in \cite{AL1}. All equations of the
ALH can be presented as the compatibility condition for the linear system

\begin{eqnarray}
\Psi_{n+1} &=& U_{n} \Psi_{n}
\label{zcr-sp}\\
{\partial \over \partial z_{j}} \Psi_{n} &=& V_{n}^{(j)} \Psi_{n},
\qquad
j = \pm 1, \pm 2, \dots
\label{zcr-evol}
\end{eqnarray}
where $\Psi_{n}$ is a $2$-column, $U_{n}$ and  $V_{n}$ are $2 \times 2$
matrices with $U_{n}$ being given by

\begin{equation}
U_{n} = U_{n} (\lambda) =
\pmatrix{ \lambda & r_{n} \cr q_{n} & \lambda^{-1} }
\end{equation}
(here $\lambda$ is the auxiliary (spectral) constant parameter) and
the matrices $V_{n}^{(j)}$ are  polynomials in $\lambda$, $\lambda^{-1}$.
The ALH in a natural way can be splitted into two subsystems
(subhierarchies). One of them corresponds to the case when
$ V_{n}^{(j)}, \quad j = 1, 2, \dots$ are $j$th order polynomials in
$\lambda^{-1}$ ('positive' subhierarchy). Its simplest equations are

\begin{eqnarray}
{\partial q_{n} \over \partial z_{1}} &=&
 - i p_{n} q_{n+1}
\label{1q}
\\
{\partial r_{n} \over \partial z_{1}} &=&
\phantom{-} i p_{n} r_{n-1}
\label{1r}
\end{eqnarray}
where

\begin{equation}
p_{n} = 1 - q_{n}r_{n}
\end{equation}
The 'negative' subhierarchy is build up of the $V$-matrices being
polynomials in $\lambda$. Its simplest equations are

\begin{eqnarray}
{\partial q_{n} \over \partial \bar z_{1}} &=&
- i p_{n} q_{n-1}
\label{-1q}
\\
{\partial r_{n} \over \partial \bar z_{1}} &=&
\phantom{-} i p_{n}r_{n+1}
\label{-1r}
\end{eqnarray}
(I use the notation $\bar z_{j} = z_{-j}, \quad j = 1, 2, \dots$).

It has been shown in \cite{FR2} that the ALH can be presented in the form
of the functional-difference equations:

\begin{eqnarray}
q_{n}( z, \bar z ) - q_{n}( z - i [\xi], \bar z ) &=&
   \xi \Bigl[
         1 - q_{n}( z, \bar z ) r_{n}( z - i [\xi], \bar z )
   \Bigr]
   q_{n+1}( z, \bar z )
\label{fr-q}
\\
r_{n}( z, \bar z ) - r_{n}( z + i [\xi], \bar z) &=&
   \xi \Bigl[
         1 - q_{n}( z + i [\xi], \bar z ) r_{n}( z, \bar z )
   \Bigr]
   r_{n-1}( z, \bar z )
\label{fr-r}
\end{eqnarray}
for the 'positive' subhierarchy, and

\begin{eqnarray}
q_{n}( z, \bar z )  - q_{n}( z, \bar z - i [\xi^{-1}] ) &=&
   \xi^{-1} \Bigl[
         1 - q_{n}( z, \bar z ) r_{n}( z, \bar z - i [\xi^{-1}] )
   \Bigr]
   q_{n-1}( z, \bar z )
\label{fr-q-neg}
\\
r_{n}( z, \bar z ) - r_{n}( z, \bar z + i [\xi^{-1}]) &=&
   \xi^{-1} \Bigl[
         1 - q_{n}( z, \bar z + i [\xi^{-1}] ) r_{n}( z, \bar z )
   \Bigr]
   r_{n+1}( \bar z, z )
\label{fr-r-neg}
\end{eqnarray}
for the 'negative' one. Here the designations

\begin{equation}
f( z, \bar z ) =
f \left(
   z_{1}, z_{2}, z_{3}, \dots \; \bar z_{1}, \bar z_{2}, \bar z_{3}, \dots
\right)
\end{equation}
and

\begin{eqnarray}
f(z \pm [\xi], \bar z) &=&
f \left(
  z_{1} \pm \xi,
  z_{2} \pm \xi^{2}/2,
  z_{3} \pm \xi^{3}/3,
  \dots \;
  \bar z_{1},
  \bar z_{2},
  \bar z_{3},
  \dots
\right)
\\
f(z , \bar z \pm [\xi^{-1}] ) &=&
f \left(
  z_{1},
  z_{2},
  z_{3},
  \dots \;
  \bar z_{1} \pm \xi^{-1},
  \bar z_{2} \pm \xi^{-2}/2,
  \bar z_{3} \pm \xi^{-3}/3,
  \dots
\right)
\end{eqnarray}
are used. Expanding equations (\ref{fr-q}), (\ref{fr-r}) in power series in
$\xi$ one can obtain all equations of the 'positive' subhierarchy.
Analogously, expanding equations (\ref{fr-q-neg}), (\ref{fr-r-neg}) in
power series in $\xi^{-1}$ one can obtain all equations of the 'negative'
one.

In what follows I will use also the tau-functions of the ALH,
$\sigma_{n}$, $\rho_{n}$ and $\tau_{n}$, which are defined by

\begin{equation}
q_{n} = { \sigma_{n} \over \tau_{n} },
\qquad
r_{n} = { \rho_{n} \over \tau_{n} }
\label{tau-def}
\end{equation}
and

\begin{equation}
\tau_{n-1} \tau_{n+1} = \tau_{n}^{2} - \sigma_{n} \rho_{n}
\label{tau-pqr}
\end{equation}
The functional representation of the ALH in terms of the tau-functions can
be written as

\begin{eqnarray}
\tau_{n}  ( z ) \; \sigma_{n}( z + i[\xi] ) -
\sigma_{n}( z ) \; \tau_{n}  ( z + i[\xi] ) &=&
\xi \;
\tau_{n-1}( z ) \; \sigma_{n+1}( z + i[\xi] )
\label{fr-sigma}
\\
\rho_{n}  ( z ) \; \tau_{n}  ( z + i[\xi] ) -
\tau_{n}  ( z ) \; \rho_{n}  ( z + i[\xi] ) &=&
\xi  \;
\rho_{n-1}( z ) \; \tau_{n+1}( z + i[\xi] )
\label{fr-rho}
\\
\tau_{n}  ( z ) \; \tau_{n}  ( z + i[\xi] ) -
\rho_{n}  ( z ) \; \sigma_{n}( z + i[\xi] ) &=&
\phantom{\xi} \;
\tau_{n-1}( z ) \; \tau_{n+1}( z + i[\xi] )
\label{fr-tau}
\end{eqnarray}
(where dependence on $\bar z_{j}$'s is omitted) and

\begin{eqnarray}
\tau_{n}  ( \bar z ) \; \sigma_{n}( \bar z + i[\xi^{-1}] ) -
\sigma_{n}( \bar z ) \; \tau_{n}  ( \bar z + i[\xi^{-1}] ) &=&
\xi^{-1} \;
\tau_{n+1}( \bar z ) \; \sigma_{n-1}( \bar z + i[\xi^{-1}] )
\label{fr-sigma-neg}
\\
\rho_{n}  ( \bar z ) \; \tau_{n}  ( \bar z + i[\xi^{-1}] ) -
\tau_{n}  ( \bar z ) \; \rho_{n}  ( \bar z + i[\xi^{-1}] ) &=&
\xi^{-1}  \;
\rho_{n+1}( \bar z ) \; \tau_{n-1}( \bar z + i[\xi^{-1}] )
\label{fr-rho-neg}
\\
\tau_{n}  ( \bar z ) \; \tau_{n}  ( \bar z + i[\xi^{-1}] ) -
\rho_{n}  ( \bar z ) \; \sigma_{n}( \bar z + i[\xi^{-1}] ) &=&
\phantom{\xi^{-1}} \;
\tau_{n+1}( \bar z ) \; \tau_{n-1}( \bar z + i[\xi^{-1}] )
\label{fr-tau-neg}
\end{eqnarray}
(where dependence on $z_{j}$'s is omitted).

The key idea of the present work is to establish the relation between
these equations and the famous Fay's identity for the $\theta$-functions
which can be used to derive the finite-gap QPS of the ALH.

%%%%%%%%%%%%%%%%%%%%%%%%%%%%%%%%%%%%%%%%%%%%%%%%%%%%%%
\section{Fay's identity.}              \label{sec_Fay}
%%%%%%%%%%%%%%%%%%%%%%%%%%%%%%%%%%%%%%%%%%%%%%%%%%%%%%
\setcounter{equation}{0}

In this paper we will deal with the compact Riemann surface $X$ of the
genus $g$ corresponding to the hyperelliptic curve

\begin{equation}
s^{2} = \mathcal{P}_{2g+2}(\xi)
\end{equation}
where $\mathcal{P}_{2g+2}(\xi)$ is a polynomial without multiple roots of
degree $2g+2$.  In the framework of the IST such
curves appear in the analysis of the scattering problem (\ref{zcr-sp}).
For example, in the case of the periodic conditions

\begin{equation}
q_{n+g+1} = q_{n},
\qquad
r_{n+g+1} = r_{n}
\label{per-cond}
\end{equation}
the polynomial $\mathcal{P}_{2g+2}(\xi)$ is defined by

\begin{equation}
\mathcal{P}_{2g+2}(\lambda^{2}) =
\lambda^{2(g+1)} \left\{
\left[ {\rm tr}\;T_{n}(\lambda) \right]^{2} - 4 \det T_{n}(\lambda)
\right\}
\label{poly}
\end{equation}
where $T_{n}(\lambda)$ is the transfer matrix of the scattering problem
(\ref{zcr-sp}),

\begin{equation}
T_{n}(\lambda) = U_{n+g}(\lambda) \dots U_{n}(\lambda)
\end{equation}
(it can be straightforwardly shown that the right-hand side of (\ref{poly})
under the restriction (\ref{per-cond}) does not depend on the index $n$).
Topologically, $X$ is a sphere with $g$ handles. One can
choose a set of $2g$ closed contours (cycles) 
$\{ a_{i}, b_{i} \}_{i=1, ..., g}$ with the intersection indices

\begin{equation}
a_{i} \circ a_{j} =
b_{i} \circ b_{j} = 0,
\qquad
a_{i} \circ b_{j} = \delta_{ij}
\qquad
i,j = 1, \dots, g
\end{equation}
and find $g$ independent holomorphic differentials, say ones given locally 
by

\begin{equation}
\widetilde{\omega}_{k} =
{ \xi^{k-1} d \xi \over \sqrt{\mathcal{P}_{2g+2}(\xi)} },
\qquad
k=1, \dots, g
\end{equation}
which can be used to construct the canonical basis of the holomorphic
1-forms

\begin{equation}
\omega_{k} = \sum_{l=1}^{g} C_{k,l} \widetilde{\omega_{l}}
\end{equation}
where $\omega_{k}$'s satisfy the normalization conditions

\begin{equation}
\oint_{a_{i}} \omega_{k} = \delta_{ik}
\end{equation}
Then, the matrix of the $b$-periods,

\begin{equation}
\Omega_{ik} = \oint_{b_{i}} \omega_{k}
\end{equation}
determines the so-called period lattice,
$L_{\Omega} = \left\{
   \vec{m} + \Omega \vec{n},
   \quad
   \vec{m}, \vec{n} \in {\mathbb Z}^{g}
   \right\}$,
the Jacobian of this surface
$\mathrm{Jac}(X)={\mathbb C}^{g}/L_{\Omega}$ (2$g$ torus) and the Abel
mapping $X \to \mathrm{Jac}(X)$,

\begin{equation}
P \to \int^{P}_{P_{0}} \vomega
\label{abel}
\end{equation}
where $\vomega$ is the $g$-vector of the 1-forms,
$\vomega =
\left( \omega_{1}, \dots, \omega_{g} \right)^{\scriptscriptstyle T}$
and $P_{0}$ is some fixed point of $X$.

A central object of the theory of the compact Riemann surfaces is the
$\theta$-function, $\theta(\vzeta)=\theta(\vzeta,\Omega)$,

\begin{equation}
\theta\left(\vzeta\right) =
\sum_{ \vec{n} \, \in \, {\mathbb Z}^{g} }
\exp\left\{
   \pi i \, \vec{n} \Omega \vec{n} \; +
   2 \pi i \, \vec{n} \vzeta
\right\}
\end{equation}
which is a quasiperiodic function on $\mathbb{C}^{g}$

\begin{eqnarray}
\theta\left(\vzeta + \vec{n}\right) &=&
   \theta\left(\vzeta\right)
\\
\theta\left(\vzeta + \Omega\vec{n}\right) &=&
   \exp\left\{
      - \pi i \, \vec{n} \Omega \vec{n} \;
      - 2 \pi i \, \vec{n} \vzeta
   \right\}
\theta\left(\vzeta\right)
\end{eqnarray}
for $\vec{n} \in {\mathbb Z}^{g}$.

To simplify the following formulae I will use the designations

\begin{equation}
\th{A}{B}{\vzeta} = \theta\left( \vzeta + \int^{A}_{B} \vomega \right)
\end{equation}
and

\begin{equation}
\pf{A}{B} =
\theta
\left[ \vdelta \right]
\left( \int^{A}_{B} \vomega \right)
\label{pf}
\end{equation}
Here
$\theta \left[ \vec{c} \right] \left( \vzeta \right)$
is the so-called $\theta$-function with characteristics,

\begin{equation}
\theta \left[ \vec{c} \right]
\left( \vzeta \right) =
\exp\left\{
   \pi i \, \vec{a} \Omega \vec{a} \; +
   2 \pi i \, \vec{a} \left(\vzeta + \vec{b}\right)
\right\}
\theta \left( \vzeta + \Omega \vec{a} + \vec{b}\right),
\qquad
\vec{c} = \left( \vec{a}, \vec{b} \right)
\end{equation}
and
$\vdelta=\left( \vdelta', \vdelta'' \right)
\in {1 \over 2} \mathbb{Z}^{2g} / \mathbb{Z}^{2g}$ is a non-singular odd
characteristics,

\begin{equation}
\theta \left[ \vdelta \right]
\left( \vec{0} \right) = 0,
\qquad
\mathrm{grad}_{\vzeta} \;
\theta \left[ \vdelta \right]
\left( \vec{0} \right) \ne \vec{0}
\end{equation}
Function (\ref{pf}) is closely related to the prime form
\cite{Fay,Mumford},

\begin{equation}
E(P,Q) =
{ \pf{P}{Q}
  \over
  \sqrt{\chi(P)} \sqrt{\chi(Q)}
}
\end{equation}
where $\chi$ is given by

\begin{equation}
\chi(P) =
\sum_{i=1}^{g}
\left(
  {\partial \over \partial \zeta_{i} } \;
  \theta \left[ \vdelta \right]
\right) \left( \vec{0} \right)
  \omega_{i}( P )
\end{equation}
The prime form $E(P,Q)$ is skew-symmetric, $E(P,Q) = - E(Q,P)$, has
first-order zero along the diagonal $P=Q$ and is otherwise non-zero.
Analogously,

\begin{equation}
\pf{P}{Q} = - \pf{Q}{P},
\qquad
\pf{P}{P} = 0
\end{equation}

One of the most interesting results of the theory of the $\theta$-functions
is the following identity for the $\theta$-functions associated with the
finite-genus Riemann surfaces, the Fay's identity:

\begin{equation}
\pf{P_{1}}{P_{3}} \pf{P_{4}}{P_{2}}
   \th{P_{1}}{P_{2}}{\vzeta} \th{P_{4}}{P_{3}}{\vzeta} -
\pf{P_{1}}{P_{2}} \pf{P_{4}}{P_{3}}
   \th{P_{1}}{P_{3}}{\vzeta} \th{P_{4}}{P_{2}}{\vzeta} =
\pf{P_{2}}{P_{3}} \pf{P_{4}}{P_{1}}
   \th{}{}{\vzeta}   \th{P_{1}P_{4}}{P_{2}P_{3}}{\vzeta}
\label{fay}
\end{equation}
(here $P_{1}, ..., P_{4}$ are arbitrary points of $X$) and namely this
formula will be basis of the following consideration.

%%%%%%%%%%%%%%%%%%%%%%%%%%%%%%%%%%%%%%%%%%%%%%%%%%%%%%
\section{Quasiperiodic solutions.}   \label{sec_QPS}
%%%%%%%%%%%%%%%%%%%%%%%%%%%%%%%%%%%%%%%%%%%%%%%%%%%%%%
\setcounter{equation}{0}

It is already known that in the quasiperiodic case the tau-functions of
the ALH are (up to some simple factors) the $\theta$-functions of
different arguments and I am going now to present the Fay's identity and
some of its corollaries in the form similar to
(\ref{fr-sigma})--(\ref{fr-tau}) and
(\ref{fr-sigma-neg})--(\ref{fr-tau-neg})
which will enable to obtain the finite-gap solutions of these functional
equations, i.e.  to obtain the finite-gap solutions of the ALH.

Hereafter I will use the letters $A$, $B$, $C$, $D$ for the points of the
Riemann surface which correspond to the points $0$ and $\infty$ of the
complex plane,

\begin{equation}
A = \infty_{+}, \quad
D = \infty_{-}, \quad
B = 0_{-},      \quad
C = 0_{+}
\end{equation}
Since $A,D$ and $B,C$ are poles and zeroes of the meromorphic function
$\pi(P)$, which is a projection of $X$ onto the extended complex plane
$\mathbb{P}^{1}$ sending a point $P=(s,\xi)$ into $\xi$, they satisfy,
according to the Abel's theorem, the condition
$\int^{AD}_{BC} \vomega \in L_{\Omega}$. The integration pathes in
(\ref{abel}) can be chosen in such a way that

\begin{equation}
\int^{AD}_{BC} \vomega = \vec{0}
\label{int-adbc}
\end{equation}
(here zero stands for $\vec{0}$ from $\mathbb{C}^{g}$, not from
$\mathrm{Jac}(X)$) and in what follows I will accept (\ref{int-adbc}) as
true.

Now I am going to use (\ref{fay}) thinking of three points from
$(P_{1}, P_{2}, P_{3}, P_{4})$ as constant (I will choose them from the
set $(A,B,C,D)$) and the fourth one (I will denote it by $P$) as variable.
Setting $(P_{1}, P_{2}, P_{3}, P_{4}) = (A, B, C, P)$ one can rewrite
(\ref{fay}) as

\begin{equation}
\pf{B}{C} \pf{P}{A} \th{}{}  {\vzeta} \th{P}{D}{\vzeta} +
\pf{A}{B} \pf{P}{C} \th{A}{C}{\vzeta} \th{P}{B}{\vzeta} =
\pf{A}{C} \pf{P}{B} \th{A}{B}{\vzeta} \th{P}{C}{\vzeta}
\label{fay-sigma}
\end{equation}
This formula is the quasiperiodic analog of (\ref{fr-sigma}).
Shifting the arguments of the $\theta$-functions,
$\vzeta \to \vzeta + \int^{C}_{A} \vomega$,
one can obtain the equation which will be transformed below to
(\ref{fr-rho}):

\begin{equation}
\pf{B}{C} \pf{P}{A} \th{C}{A}{\vzeta} \th{P}{B}{\vzeta} +
\pf{A}{B} \pf{P}{C} \th{}{}  {\vzeta} \th{CP}{AB}{\vzeta} =
\pf{A}{C} \pf{P}{B} \th{C}{B}{\vzeta} \th{P}{A}{\vzeta}
\label{fay-rho}
\end{equation}
At last, replacing in (\ref{fay})
$(P_{1}, P_{2}, P_{3}, P_{4})$ with $(A, P, C, D)$
using (\ref{int-adbc}) and making the shift
$\vzeta \to \vzeta + \int^{C}_{A} \vomega$
one can write the identity

\begin{equation}
\pf{A}{D} \pf{P}{C} \th{}{}  {\vzeta} \th{P}{B}{\vzeta} -
\pf{A}{B} \pf{P}{A} \th{C}{A}{\vzeta} \th{P}{D}{\vzeta} =
\pf{A}{C} \pf{P}{D} \th{A}{B}{\vzeta} \th{P}{A}{\vzeta}
\label{fay-tau}
\end{equation}
which is a quasiperiodic analogue of (\ref{fr-tau}).

Our first goal is to present equations (\ref{fay-sigma})--(\ref{fay-tau})
in the bilinear form. To this end I will first shift the arguments of the
$\theta$-functions: $\vzeta \to \vzeta_{n}$,

\begin{equation}
\vzeta_{n} = \vzeta + n \int^{B}_{A} \vomega
\end{equation}
Next, I will introduce the functions $\sigma_{n}(P)$, $\rho_{n}(P)$ and
$\tau_{n}(P)$,

\begin{eqnarray}
\tau_{n}(P)   &=& \alpha_{n}(P) \; \th{P}{B}{\vzeta_{n}}
\\
\sigma_{n}(P) &=& \beta_{n}(P)  \; \th{P}{D}{\vzeta_{n}}
\\
\rho_{n}(P)   &=& \gamma_{n}(P) \; \th{CP}{AB}{\vzeta_{n}}
\end{eqnarray}
It is not difficult to verify that if one chooses the functions
$\alpha_{n}$, $\beta_{n}$, $\gamma_{n}$ as follows

\begin{eqnarray}
\alpha_{n}(P) &=&
   \alpha_{*}
   \mu^{ n^{2} / 2 }
   \exp\left\{ n \varphi_{\scriptscriptstyle DC} ( P ) \right\}
\\
\beta_{n}(P) &=&
   q_{*}
   \varepsilon^{n}
   \exp\left\{ \varphi_{\scriptscriptstyle AC} ( P ) \right\}
   \alpha_{n}(P)
\\
\gamma_{n}(P) &=&
   r_{*}
   \varepsilon^{-n}
   \exp\left\{ - \varphi_{\scriptscriptstyle AC} ( P ) \right\}
   \alpha_{n}(P)
\end{eqnarray}
where the functions $\varphi_{\scriptscriptstyle QR}$ are defined in the
vicinity of the point $B$ by

\begin{equation}
\exp\left\{
   \varphi_{\scriptscriptstyle QR} ( P ) -
   \varphi_{\scriptscriptstyle QR} ( B )
   \right\} =
{\pf{P}{Q} \over \pf{P}{R}}
{\pf{B}{R} \over \pf{B}{Q}},
\end{equation}
the constant $\mu$ is given by

\begin{equation}
\mu = { \left( \pf{A}{C} \right)^{2} \over \pf{A}{D} \pf{B}{C} }
\label{const-mu}
\end{equation}
and $\alpha_{*}$, $q_{*}$, $r_{*}$ and $\varepsilon$ are
arbitrary constants satisfying

\begin{equation}
q_{*} r_{*} =
- { \left( \pf{A}{B} \right)^{2} \over \pf{A}{D} \pf{B}{C} }
\label{const-qr}
\end{equation}
then (\ref{fay-sigma})--(\ref{fay-tau}) can be
rewritten in terms of the functions $\sigma_{n}(P)$, $\rho_{n}(P)$ and
$\tau_{n}(P)$ as

\begin{eqnarray}
\tau_{n}  (B) \; \sigma_{n}(P) -
\sigma_{n}(B) \; \tau_{n}  (P)
&=&
K(P) \;
\tau_{n-1}(B) \; \sigma_{n+1}(P)
\label{fay-sigma-1}
\\
\rho_{n}(B) \; \tau_{n}(P) -
\tau_{n}(B) \; \rho_{n}(P)
&=&
K(P) \;
\rho_{n-1}(B) \; \tau_{n+1}(P)
\label{fay-rho-1}
\\
\tau_{n}(B) \; \tau_{n}(P) -
\rho_{n}(B) \; \sigma_{n}(P)
&=&
\phantom{K(P)} \;
\tau_{n-1}(B) \; \tau_{n+1}(P)
\label{fay-tau-1}
\end{eqnarray}
where

\begin{equation}
K(P) =
{1 \over \varepsilon} \;
{ \pf{A}{D} \over \pf{B}{C} } \;
{ \pf{P}{B} \pf{P}{C} \over \pf{P}{A} \pf{P}{D}  }
\end{equation}
Thus we have presented the Fay' identities in the bilinear form similar to
(\ref{fr-sigma})--(\ref{fr-tau}). What I have to do now is to introduce
a $z$-dependence in such a way that a shift over the Riemann surface from
the point $B$ to a point $P$ (which correspond to the points $0$ and $\xi$
of the complex plane) can be taken into account by the simultaneous shifts
$z_{m} \to z_{m} + i \xi^{m} / m$:

\begin{eqnarray}
f_{n}(B) &=&
   f_{n}(z) = f_{n}(z_{1}, z_{2}, z_{3}, ... ),
\\
f_{n}(P) &=&
   f_{n}( z + i [\xi] ) =
   f_{n}(z_{1} + i \xi, z_{2} + i \xi^{2}/2, z_{3} + i \xi^{3}/3, ... )
\end{eqnarray}
(I hope that the usage of the same letters for functional dependence on
both the point of the Riemann surface and the ALH variables $z_{m}$ will
not lead to confusion). In other words I want to introduce such functions
$\vzeta(z_{1}, z_{2}, ... )$ and
$\varphi_{\scriptscriptstyle QR}(z_{1}, z_{2}, ... )$ that

\begin{equation}
\vzeta( z + i [\xi] ) - \vzeta( z ) = \int^{P}_{B} \vomega
\label{shift-zeta}
\end{equation}
and

\begin{equation}
\varphi_{\scriptscriptstyle QR} ( z + i [\xi] ) -
\varphi_{\scriptscriptstyle QR} ( z ) =
\varphi_{\scriptscriptstyle QR} ( P ) -
\varphi_{\scriptscriptstyle QR} ( B )
\label{shift-varphi}
\end{equation}
This can be done as follows.  In the neighborhood of the point $B$ (which
is an preimage of the point $\xi=0$ of the complex plane) the components
of the integral in (\ref{shift-zeta}) can be presented in terms of the
$\xi$-coordinate as

\begin{equation}
\int_{B}^{P} \omega_{k} =
W_{k}(\xi) =
- \sum_{l=1}^{g}
  C_{k,l}
  \int_{0}^{\xi}
  { x^{l-1} \mathrm{d} x \over \sqrt{{\cal P}_{2g+2}(x)} }
\label{W-k}
\end{equation}
where the sign of the square root is fixed by $\sqrt{1}=1$.
Hence, taking $\vzeta$ to be a linear function of
the coordinates $z_{m}$,

\begin{equation}
\vzeta = \vzeta(z) = \sum_{m=1}^{\infty} \vzeta^{(m)} \; z_{m}
\end{equation}
one can conclude that to satisfy (\ref{shift-zeta}) the vectors
$\vzeta^{(m)}$ should be defined as the coefficients of the series

\begin{equation}
\sum_{m=1}^{\infty} \vzeta^{(m)} \; \xi^{m} =
 - i \, \xi { \mathrm{d} \over \mathrm{d}\xi} \; \vec{W}(\xi)
\label{zeta-m}
\end{equation}
(here $\vec{W}$ is the vector with the components $W_{k}$).
Using (\ref{W-k}) one can rewrite (\ref{zeta-m}) as

\begin{equation}
\sum_{m=1}^{\infty} \zeta_{k}^{(m)} \; \xi^{m} =
i \sum_{l=1}^{g}
  C_{k,l}
  { \xi^{l} \over \sqrt{{\cal P}_{2g+2}(\xi)} }
\end{equation}
(the right-hand side of this equation should be understood as a power
series in $\xi$).

In a similar way one can tackle equation (\ref{shift-varphi}) and to
derive the following result: $\varphi_{\scriptscriptstyle QR} ( z )$ is
the linear function,

\begin{equation}
\varphi_{\scriptscriptstyle QR} ( z ) =
\sum_{m=1}^{\infty}
\varphi_{\scriptscriptstyle QR}^{(m)} \; z_{m}
\label{phi-taylor}
\end{equation}
with the coefficients $\varphi_{\scriptscriptstyle QR}^{(m)}$ being
defined by

\begin{equation}
\sum_{m=1}^{\infty}
\varphi_{\scriptscriptstyle QR}^{(m)} \; \xi^{m} =
 - i \xi { \mathrm{d} \over \mathrm{d}\xi} \;
\ln{
   \theta\left[ \vdelta \right]
   \left( \int^{B}_{Q} \vomega + \vec{W}(\xi) \right)
\over
   \theta\left[ \vdelta \right]
   \left( \int^{B}_{R} \vomega + \vec{W}(\xi) \right)
}
\end{equation}
(the right-hand side is again a power series in $\xi$).

Thus one can write the following expressions for the tau-functions:

\begin{eqnarray}
\tau_{n}( z ) &=&
   \alpha_{*}
   \mu^{ n^{2} / 2 }
   \exp\left\{
      \; n \; \varphi_{\scriptscriptstyle DC} ( z )
   \right\} \;
   \theta\left( \vzeta_{n}( z ) \right)
\label{pos-tau}
\\
\sigma_{n}( z ) &=&
   q_{*} \varepsilon^{n}
   \mu^{ n^{2} / 2 }
   \exp\left\{
      \; n \; \varphi_{\scriptscriptstyle DC} ( z ) +
      \varphi_{\scriptscriptstyle AC} ( z )
   \right\} \;
   \theta\left( \vzeta_{n}( z ) - \int^{C}_{A} \vomega \right)
\label{pos-sigma}
\\
\rho_{n}( z ) &=&
   r_{*} \varepsilon^{-n}
   \mu^{ n^{2} / 2 }
   \exp\left\{
      \; n \; \varphi_{\scriptscriptstyle DC} ( z ) -
      \varphi_{\scriptscriptstyle AC} ( z )
   \right\} \;
   \theta\left( \vzeta_{n}( z ) + \int^{C}_{A} \vomega \right)
\label{pos-rho}
\end{eqnarray}

At last, we have to rewrite the function $K(P)$ from the right hand side
of (\ref{fay-sigma-1})--(\ref{fay-tau-1}). This is the first time, since
the Fay's identity has been written down, that we need some facts from the
theory of the Riemann surfaces -- till now all was done by simple algebra.
Consider the function

\begin{equation}
f(P) = { \pf{P}{B} \pf{P}{C} \over \pf{P}{A} \pf{P}{D}  }
\end{equation}
This is a single-valued, due to the condition (\ref{int-adbc}), function
which possesses zeroes at the points $B$, $C$ and poles at $A$, $D$.
Remembering that $B$, $C$ correspond to $\xi=0$, and $A$, $D$ -- to
$\xi=\infty$, one can easily obtain one function with the same divisor,
$B+C-A-D$, namely, the projection $\pi(P)$ discussed above (see paragraph
before (\ref{int-adbc})). The quotient $\pi(P)/f(P)$ has no poles (and
zeroes as well) on $X$, hence it is a constant

\begin{equation}
f(P) = C \xi \qquad \mbox{for} \qquad P=(s,\xi)
\end{equation}
Thus, if we take

\begin{equation}
\varepsilon = C \, { \pf{A}{D} \over \pf{B}{C} }
\label{const-eps}
\end{equation}
then

\begin{equation}
K(P) = \xi
\end{equation}
and relations (\ref{fay-sigma-1})--(\ref{fay-tau-1}) become
(\ref{fr-sigma})--(\ref{fr-tau}), or in other words, the functions defined
by (\ref{pos-tau})--(\ref{pos-rho}) solve equations
(\ref{fr-sigma})--(\ref{fr-tau}).

Till now we were operating in a neighborhood of the point $B$ and
obtained solutions of equations (\ref{fr-sigma})--(\ref{fr-tau}), and
hence of (\ref{fr-q})--(\ref{fr-r}), i.e. solved the 'positive' part of
the ALH. To take into account the 'negative' equations
(\ref{fr-q-neg})--(\ref{fr-r-neg}), or
(\ref{fr-sigma-neg})--(\ref{fr-tau-neg}), one can proceed in the similar
way, but this time considering flows near another distinguished point,
$D$, which is an preimage of the point $\xi=\infty$. It can be shown that
functions $\tau_{n}$, $\sigma_{n}$, $\rho_{n}$ given by
(\ref{pos-tau})--(\ref{pos-rho}) will solve
(\ref{fr-sigma-neg})--(\ref{fr-tau-neg}) provided we introduce the
$\bar z$-dependence by replacing

\begin{eqnarray}
\vzeta(z) & \to & \vzeta(z, \bar z)
\\
\varphi_{\scriptscriptstyle DC} ( z )  & \to &
\varphi_{\scriptscriptstyle DC} ( z ) +
\bar\varphi_{\scriptscriptstyle BA} ( \bar z )
\\
\varphi_{\scriptscriptstyle AC} ( z )  & \to &
\varphi_{\scriptscriptstyle AC} ( z ) +
\bar\varphi_{\scriptscriptstyle CA} ( \bar z )
\end{eqnarray}
(the overbar {\em does not} mean the complex conjugation!)
where

\begin{equation}
\vzeta( z, \bar z + i [\xi^{-1}] ) -
\vzeta( z, \bar z ) =
\overline{\vec{W}}(\xi^{-1}) =
\int^{P}_{D} \vomega
\end{equation}
and

\begin{equation}
\bar\varphi_{\scriptscriptstyle QR} ( \bar z + i [\xi^{-1}] ) -
\bar\varphi_{\scriptscriptstyle QR} ( \bar z ) =
\ln { \pf{P}{Q} \over \pf{P}{R} } \; { \pf{D}{R} \over \pf{D}{Q} }
\end{equation}
Thus, now we have all necessary to formulate the main result of this
paper. The finite genus solutions of the ALH can be presented as

\begin{eqnarray}
q_{n}( z, \bar z ) &=&
   q_{*} \varepsilon^{n}
   \exp\left\{   \varphi(z,\bar z) \right\} \;
   { \theta\left( \vzeta(z,\bar z) + n\vec{U} - \vec{V} \right)
	\over
	\theta\left( \vzeta(z,\bar z) + n\vec{U} \right)
   }
\label{qps-q}
\\
r_{n}( z, \bar z ) &=&
   r_{*} \varepsilon^{-n}
   \exp\left\{ - \varphi(z,\bar z) \right\} \;
   { \theta\left( \vzeta(z,\bar z) + n\vec{U} + \vec{V} \right)
	\over
	\theta\left( \vzeta(z,\bar z) + n\vec{U} \right)
   }
\label{qps-r}
\end{eqnarray}
where

\begin{equation}
\vec{U} = \int^{B}_{A} \vomega,
\qquad
\vec{V} = \int^{C}_{A} \vomega
\end{equation}
The functions $\vzeta( z, \bar z )$ and $\varphi( z, \bar z )$ are given by

\begin{eqnarray}
\vzeta( z, \bar z ) &=&
   \sum_{m=1}^{\infty}
   \left(
   \vzeta^{(m)} \; z_{m} +
   \bar{\vzeta}^{(m)} \; \bar z_{m}
   \right) +
   \mbox{constant}
\\
\varphi( z, \bar z ) &=&
   \sum_{m=1}^{\infty}
   \left(
   \varphi^{(m)} \; z_{m} +
   \bar\varphi^{(m)} \; \bar z_{m}
   \right) +
   \mbox{constant}
\end{eqnarray}
where the constants
$\vzeta^{(m)}$, $\bar{\vzeta}^{(m)}$ and
$\varphi^{(m)}$, $\bar\varphi^{(m)}$ are defined as coefficients of the
series

\begin{eqnarray}
\sum_{m=1}^{\infty} \zeta_{k}^{(m)} \; \xi^{m} &=&
  \phantom{-}
  i \sum_{l=1}^{g}
  C_{k,l}
  { \xi^{l} \over \sqrt{ \mathcal{P}_{2g+2}(\xi) } }
\\
\sum_{m=1}^{\infty} \bar{\zeta}_{k}^{(m)} \; \xi^{-m} &=&
  -i \sum_{l=1}^{g}
  C_{k,g+1-l}
  { \xi^{-l} \over \sqrt{\overline{\mathcal{P}}_{2g+2}(1/\xi)} }
\end{eqnarray}
with

\begin{equation}
\overline{\mathcal{P}}_{2g+2}(\xi) =
\xi^{2g+2}(\xi)
\mathcal{P}_{2g+2}(1/\xi)
\end{equation}
and

\begin{eqnarray}
\sum_{m=1}^{\infty}
\varphi^{(m)} \; \xi^{m} &=&
i \, \xi { \mathrm{d} \over \mathrm{d}\xi} \;
\ln{
   \theta\left[ \vdelta \right]
   \left( \vec{U} - \vec{V} + \vec{W}(\xi) \right)
   \over
   \theta\left[ \vdelta \right]
   \left( \vec{U} + \vec{W}(\xi) \right)
   }
\\
\sum_{m=1}^{\infty}
\bar\varphi^{(m)} \; \xi^{-m}  &=&
i \, \xi { \mathrm{d} \over \mathrm{d}\xi} \;
\ln{
   \theta\left[ \vdelta \right]
   \left( \vec{U} + \overline{\vec{W}}(1/\xi) \right)
   \over
   \theta\left[ \vdelta \right]
   \left( \vec{U} + \vec{V} + \overline{\vec{W}}(1/\xi) \right)
   }
\end{eqnarray}
The constant $\varepsilon$ is given by (\ref{const-eps}) and $q_{*}$,
$r_{*}$ are arbitrary constants related by (\ref{const-qr}).

The 'real' tau-function $\tau_{n}$ can be written as

\begin{equation}
\tau_{n}( z, \bar z ) =
   \alpha_{*}
   \mu^{ n^{2} / 2 }
   \exp\left\{ \; n \; \psi ( z, \bar z ) \right\} \;
   \theta\left( \vzeta(z,\bar z) + n\vec{U} \right)
\end{equation}
where the constant $\mu$ is given by (\ref{const-mu}),

\begin{equation}
\psi( z, \bar z ) =
   \sum_{m=1}^{\infty}
   \left(
   \psi^{(m)} \; z_{m} +
   \bar\psi^{(m)} \; \bar z_{m}
   \right) +
   \mbox{constant}
\end{equation}
and

\begin{eqnarray}
\sum_{m=1}^{\infty}
\psi^{(m)} \; \xi^{m} &=&
i \, \xi { \mathrm{d} \over \mathrm{d}\xi} \;
\ln{
   \theta\left[ \vdelta \right]
   \left( \vec{U} - \vec{V} + \vec{W}(\xi) \right)
   \over
   \theta\left[ \vdelta \right]
   \left( -\vec{V} + \vec{W}(\xi) \right)
   }
\\
\sum_{m=1}^{\infty}
\bar\psi^{(m)} \; \xi^{-m}  &=&
i \, \xi { \mathrm{d} \over \mathrm{d}\xi} \;
\ln{
   \theta\left[ \vdelta \right]
   \left( \vec{V} + \overline{\vec{W}}(1/\xi) \right)
   \over
   \theta\left[ \vdelta \right]
   \left( \vec{U} + \vec{V} + \overline{\vec{W}}(1/\xi) \right)
   }
\end{eqnarray}

%%%%%%%%%%%%%%%%%%%%%%%%%%%%%%%%%%%%%%%%%%%%%%%%%%%%%%
\section{Discussion.}    \label{sec_D}
%%%%%%%%%%%%%%%%%%%%%%%%%%%%%%%%%%%%%%%%%%%%%%%%%%%%%%
\setcounter{equation}{0}

In this paper we have obtained the finite genus solutions for the ALH. The
results can be used to derive also the finite genus solutions for
other integrable hierarchies, which can be 'embedded' into the ALH (see
\cite{2DTL,DS,FR2,ICTP}. So, for example, the functions

\begin{eqnarray}
Q &=& { \sigma_{1} \over \tau_{0} } =
  Q_{*}
  \exp\left\{ \psi(z,\bar z) + \varphi(z,\bar z) \right\} \;
  { \theta\left( \vzeta(z,\bar z) - \widetilde{\vec{V}} \right)
    \over
    \theta\left( \vzeta(z,\bar z) \right)
   }
\\
R &=& { \rho_{-1} \over \tau_{0} } =
  R_{*}
  \exp\left\{ -\psi(z,\bar z) - \varphi(z,\bar z) \right\} \;
  { \theta\left( \vzeta(z,\bar z) + \widetilde{\vec{V}} \right)
    \over
    \theta\left( \vzeta(z,\bar z) \right)
   }
\end{eqnarray}
where

\begin{equation}
\widetilde{\vec{V}} =
\vec{V} - \vec{U} =
\int^{C}_{B} \vomega,
\end{equation}
and constants $Q_{*}$, $R_{*}$ are related by

\begin{equation}
Q_{*} R_{*} =
- \left[
   \varepsilon \;
   { \pf{A}{B} \pf{A}{C} \over \pf{A}{D} \pf{B}{C} }
\right]^{2}
\end{equation}
solve the nonlinear Scr\"odinger equation

\begin{eqnarray}
i\partial_{2} Q+\partial_{11} Q + 2 Q^{2}R = 0
\\
-i\partial_{2} R+\partial_{11} R + 2QR^{2} = 0
\end{eqnarray}
where $\partial_{m} = \partial / \partial z_{m}$, as well as all higher
equations of the hierarchy (see \cite{FR2}).

The quantities

\begin{equation}
p_{n} = { \tau_{n+1} \tau_{n-1} \over \tau_{n}^{2} } =
\mu
{
   \theta\left( \vzeta_{T}(x,\bar x) + (n-1)\vec{U} \right)
   \theta\left( \vzeta_{T}(x,\bar x) + (n+1)\vec{U} \right)
\over
   \theta^{2}\left( \vzeta_{T}(x,\bar x) + n\vec{U} \right)
}
\end{equation}
where

\begin{equation}
\vzeta_{T}(x,\bar x) =
\vzeta^{(1)} x + \bar{\vzeta}^{(1)} \bar x + \mathrm{constant}
\end{equation}
solve the 2D Toda lattice equation

\begin{equation}
{\partial^{2} \over \partial x \partial \bar x} \ln p_{n}
= p_{n-1} - 2 p_{n} + p_{n+1}
\end{equation}

In \cite{DS} the relations between the ALH and the Davey-Stewartson
equation (together with the Ishimori model) have been derived. One can
find there expressions for the corresponding finite genus solutions.

The last example stems from the fact that for any fixed $n$ the quantity

\begin{equation}
u = r_{n-1}p_{n}q_{n+1} =
{ \rho_{n-1} \sigma_{n+1} \over \tau_{n}^{2} }
\label{KP-u}
\end{equation}
solves the Kadomtsev-Petviashvili (KP) equation,

\begin{equation}
\partial_{1}
\left(
4 \partial_{3} u
+ \partial_{111} u
+ 12 u \, \partial_{1} u
\right) =
3 \partial_{22} u
\end{equation}
Thus, the results of the previous section yield the following finite genus
solution for the KP:

\begin{equation}
u =
Q_{*} R_{*}
{
  \theta\left(
    \vzeta_{KP}(z_{1}, z_{2}, z_{3}) - \widetilde{\vec{V}} \right)
  \theta\left(
    \vzeta_{KP}(z_{1}, z_{2}, z_{3}) + \widetilde{\vec{V}} \right)
\over
  \theta^{2}\left( \vzeta_{KP}(z_{1}, z_{2}, z_{3}) \right)
}
\label{KP-qps}
\end{equation}

\begin{equation}
\vzeta_{KP}(z_{1}, z_{2}, z_{3}) =
\vzeta^{(1)} z_{1} +
\vzeta^{(2)} z_{2} +
\vzeta^{(3)} z_{3} +
\mathrm{constant}
\end{equation}
Here I set $n=0$ in (\ref{KP-u}) and omitted the $\bar z_{m}$-dependence
for all $m$'s as well as the dependence on $z_{m}$ for $m>3$. This
solution differs from the already known one which corresponds to
an odd-order polynomial $\mathcal{P}_{2g+1}$ and, what is crucial, has been
obtained by considering flows near the infinity ($\pi(P)=\infty$) which in
this case is a ramification (Weierstrass) point ( the point
$\xi=\infty$ has only one preimage on the Riemann surface). I cannot at
present discuss solution (\ref{KP-qps}) in detail. For example I do
not know whether it is possible to obtain from (\ref{KP-qps}) any
non-trivial {\em real} solutions. In any case, solution (\ref{KP-qps})
seems to be interesting and worth following studies.

To conclude, I want to point out the main differences between the approach
of this paper and ones used earlier \cite{AC1,AC2,BPS,BP,MEKL}. In the
IST-based methods the hyperelliptic curves appear in the analysis of the
spectral data of the scattering problem (\ref{zcr-sp}), while the
dependence on the coordinates is derived from the system (\ref{zcr-evol}).
Here we didn't use the zero-curvature representation (\ref{zcr-sp})--
(\ref{zcr-evol}) explicitly (though it is surely hidden in the functional
equations (\ref{fr-q})--(\ref{fr-r-neg})). We started with an almost
arbitrary polynomial $\mathcal{P}_{2g+2}$ (the fact that it is related to
the transfer matrix of the scattering problem (\ref{zcr-sp}) wasn't
crucial for our consideration) and obtained the $z_{j}$-, $\bar
z_{j}$-dependence {\it directly} from the equations of the ALH (and not
from the corresponding equations for the transfer matrix $T_{n}$).

As to the algebro-geometrical method, the main distinguishing point is
that the approach of the work \cite{MEKL} (and analogous works devoted to
other integrable equations) is, so to say 'global', while ours is 'local'.
The authors of \cite{MEKL} used the Baker-Akhiezer function and other
structures defined for the whole Riemann surface $X$. At the same time we
didn't use globally defined objects: each time we introduced some
functions depending on the point $P$ of the Riemann surface $X$ it was
understood that it is defined in some vicinity of some distinguished point
($B$ or $D$). We even didn't discussed the question whether our functions,
say  $\varphi_{\scriptscriptstyle QR}(P)$, are well-defined, or
single-valued, for all $P$'s, $P \in X$. All we needed is the Taylor
expansions, say (\ref{phi-taylor}), hence for our purposes it was enough
that our functions exist locally, for $P$ belonging to some (arbitrary
small) neighborhood of the point $B$ (or $D$).

At last, I would like to note that the idea to apply the Fay's identity to
differential equations is far from new. For example, in the book
\cite{Mumford} one can find few examples of how to demonstrate that
$\theta$-functions of some arguments solve the KP, KdV, sine-Gordon
equations. However in these examples the main question is how some
{\it combinations} of differential operators (flows) act on a
$\theta$-function.  To my knowledge the problem of action of these
operators taken {\it separately} hasn't been considered before. Now we
know a partial answer:  the flows near a regular (not Weierstrass) point
of a Riemann surface can be described by means of the equations of the
ALH. Such an appearance of the ALH seems to be new and rather interesting.
Combined with the results of the works \cite{FR1,FR2,2DTL,DS,ICTP} it can
be viewed as one more point indicating the 'universality' of this
hierarchy.

%%%%%%%%%%%%%%%%%%%%%%%%%%%%%

\end{document}